\documentclass{aastex631}

\usepackage{amsmath}
\usepackage{color}

\usepackage{graphicx}

\begin{document}

\title{MARVEL Analysis of the Measured High-resolution  Spectra of CO Isotopologues}

\author{Timur Grigorev}
\author{Yuhang Dai}
\author{Max Potter}
\author{Xiaoyu Xiang}
\author{Keyu Zhang}

\author{Jonathan Tennyson}
\affiliation{\centering Department of Physics and Astronomy, University College London, London, WC1E 6BT, UK}
\correspondingauthor{Jonathan Tennyson}
\email{j.tennyson@ucl.ac.uk}



\begin{abstract}

Carbon monoxide is thought to be the second most abundant molecule in the Universe. This makes observation of both
its parent isotopologue ($^{12}$C$^{16}$O) and its stable isotopologues,
$^{13}$C$^{16}$O, $^{12}$C$^{18}$O,  $^{12}$C$^{17}$O,  $^{13}$C$^{18}$O and  $^{13}$C$^{17}$O, important in variety of 
objects. Here the MARVEL (Measured Active Rotational–Vibrational Energy Levels) algorithm is used to determine precise rotational-vibrational energy levels for the five minor isotopologues of carbon monoxide in their electronic ground state. A review of 27 literature sources yields 3716, 1454, 89, 728 and 57 validated transitions for $^{13}$C$^{16}$O, $^{12}$C$^{18}$O,  $^{12}$C$^{17}$O,  $^{13}$C$^{18}$O and  $^{13}$C$^{17}$O, respectively, giving 863, 499, 33, 345 and 45 
empirically-determined, rotation-vibration energy levels, respectively.

\end{abstract}


\section{Introduction} \label{sec:intro}

Carbon monoxide is the second most abundant molecule in interstellar space after hydrogen \citep{LiBuenker,Vazquez} where it is generally thought to have an abundance of 10\textsuperscript{-4} in comparison with H\textsubscript{2} \citep{Lacy94}. Cold carbon monoxide gas emits a clear and distinct radio signal therefore, 
it is a  good measure of the distribution of molecular gas in interstellar clouds \citep{92TuHu.CO,92YaFu.CO,25KaSoMo.CO}. The use of emissions from CO have also been used to infer the temperature of the cosmic microwave background (CMB)  for different regions \citep{11NoPeSr.CO,20KlIvPe.CO,23MaGkCo.CO}.

CO is important in many other environments including  comets, planetary atmospheres, exoplanets and the photospheres of both the Sun  and cooler 
stars \citep{73Hall,Cooper1987Kirby,91FaGuSaGr,HASE2010521,14Krasnopolsky.CO,BERNATH20173}. Due to its distinctive properties and behavior, scientists have conducted numerous experiments and theories to study its spectra. For example it has been suggested that the band head structure in CO infrared spectra allows it to be used as a thermometer in the atmosphere of cool stars \citep{jt352}. Isotope effects are much easier to observe in molecular
than atomic spectra, and the abundance of CO makes it particularly suitable for studies of C and O isotopic abundances.
Studies of CO isotopologues have been conducted in hydrogen-deficient carbon Stars \citep{25MeKaKa.CO}, young stars \citep{25PiSnDe.CO} and
red giants \citep{12AbPaBu.CO}. CO isotopologues have been detected in brown dwarfs \citep{21ZhSnMo.CO,25MuDeLa.CO} and so
far CO is the only molecule for which isotopologues have been detected in exoplanets \citep{21YaSnBo.CO,23GaDeSn.CO}.
Terrestrially, studies  of CO are particularly useful for isotope analysis for paleoclimatology (the study of past climates) \citep{Griff15,Sahl25}. Studies of CO in meteorite samples also show interesting
isotope ratios \citep{04YuKu.CO}.

Spectra of interstellar
CO have been very widely observed, indeed in some cases as a proxy for H$_2$ itself. Here we are concerned with the energy levels
of isotopically substituted CO.
Spectra of various CO isotopologues  have long been observed in the interstellar medium \citep{72PeJeWi} and are still regularly
observed and their distributions mapped. There are two main drivers for this. Firstly, information on the isotopologues themselves helps to determine local isotopic abundances \citep{01BePaWoKl} and to study  isotope-dependent chemical processes such as fractionation; for example, 
\citet{09VivaBl.CO} showed that  photodissociation and chemistry of CO isotopologues
in interstellar clouds and circumstellar disks can lead to enhanced presence of the rare isotopologues in these objects; simple fractionation effects can also lead increased local isotope abundances in colder environments \citep{80LaGoCa.CO,92ShFeLa.CO,03FeLaSh.CO}.
In general study of several isotopologues can reveal unique insights into local chemistry not
available from observing on single species. Secondly, spectra of  parent $^{12}$C$^{16}$O are often
optically thick, so CO columns can only be determined by monitoring the less abundant isototopologues.

CO isotope spectra have been used for a whole variety of other studies including determining
carbon and oxygen isotopic ratios in Arcturus and Aldebaran \citep{12AbPaBu.CO} and oxygen isotope ratios in hydrogen-deficient carbon stars.
\citep{25MeKaKa.CO}. More recently CO has become means of measuring   carbon and oxygen isotope ratios in exoplanets \citep{23GaDeSn.CO,25PiSnDe.CO}
with evidence of $^{13}$C$^{16}$O rich atmospheres \citep{21YaSnBo.CO}.
\citet{91FaGuSaGr} observed lines due to  $^{13}$C$^{16}$O, $^{12}$C$^{18}$O and $^{12}$C$^{17}$O in solar spectra; we use these data below.

Because of the importance of CO for both astronomical and terrestrial applications, 
the past five decades, numerous high-resolution experiments have been conducted to determine the positions of rotation-vibration and pure-rotation transitions  of CO isotopologues; these measurements are detailed below.
Recently, \citet{jt961} conducted a comprehensive  analysis of
$^{12}$C$^{16}$O in its electronic ground state using the the Measured Active Rotational Vibrational Energy Levels (MARVEL) procedure. The vibration-rotation spectrum of  $^{12}$C$^{16}$O is particularly well studied and this work used 19,399  vibration-rotation transitions (7955 unique) to obtain  2293 empirical energy levels.
This dataset spans levels with rotational states, $J$, up to 123 and vibrational states with $v \leq 41$. In this work we complement
this MARVEL study on the parent $^{12}$C$^{16}$O isotopologue (henceforth denoted 26) with an equivalent study of each of
the stable minor isotopologues. Unsurprisingly, there are rather fewer transitions available for these less abundant species
but we are still able to recover  empirical energy levels for each  species.
In this work we assemble high resolution spectra of CO isotopologues and use them to determine the associated energy levels for each species.
These energy levels can be used to improve available line lists \citep{94Goorvich.CO,15LiGoRo.CO,22MeErSt.CO,22MeUs.CO} or, when combined with highly accurate intensities \citep{jt871,jt970,jt987}, to produce new high accuracy line lists.

\section{ Methodological Details} \label{sec:style}
\subsection{ MARVEL}

\begin{figure}[hbt!]
    \centering
    \includegraphics[width=0.6\linewidth]{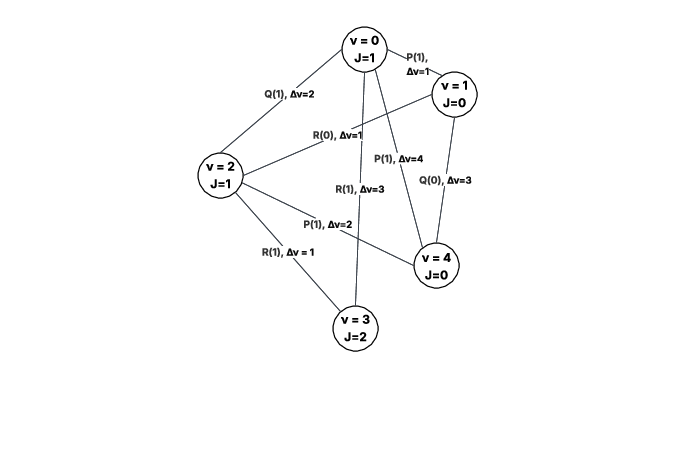}
    \caption{A simple spectroscopic network}
    \label{fig:Node}
\end{figure}

MARVEL stands for Measured Active Rotational-Vibrational Energy Levels. It forms spectroscopic networks to validate experimental transition data, resulting in the most accurate form of energy levels \citep{jt412}.
MARVEL works by forming a network with the quantum numbers representing the quantum state acting as nodes (vertices of the network) and the transitions acting as edges (connections between nodes); the edges are weighted by their uncertainties. An example of a simple spectroscopic network is shown in Figure \ref{fig:Node}.
Creating a useful network requires as much data as possible as this will allow MARVEL to create more edges and to refine the uncertainty and weighting of said edges.

To calculate the energy levels, MARVEL performs a matrix inversion based on the following equation taken from \citep{jt412}:
\begin{equation}
 aX=Y 
\end{equation}
where $Y$ is the energies vector of $N\textsubscript{\rm E}$ dimensions, $X$ is the transtions vector of $N\textsubscript{t} – 1$ dimensions, where 
$N\textsubscript{\rm E}$ is the number of energy levels (nodes) in the network and $N\textsubscript{t}$ is the number of transitions (edges). For the $i^{\rm th}$ transition matrix element $a_{ij}$ of matrix $a$ is given by:
\begin{align}
    -1, &\quad \text{if $E_j$ is the lower energy level,}\\
    +1, &\quad \text{if $E_j$ is the upper energy level,}\\
    0,  &\quad \text{otherwise}\\ 
\end{align}
The weighted least-squares solution is given by:
\begin{equation}
\textbf{AX = B}
\end{equation}
where
\begin{eqnarray*}
 A = a^T g a \\
B = a^T g Y \\
g_i = 1/\delta_i^2
\end{eqnarray*}
and $\delta_{i}$ is the uncertainty of the $i\textsuperscript{th }$ transition, $\textbf{A}$ is a $(N\textsubscript{t}-1) \times (N\textsubscript{t}-1)$ dimension matrix. 
The unweighted version of the equation is a simple system of linear equations which is much easier to solve than the overdetermined system of linear equations \citep{jt412}.

The MARVEL code \citep{jt412} requires input in the form of a specially formatted text file, where each line represents an energy transition obtained from a literature source. The first two columns contain the experimental line positions, along with their uncertainties, both expressed in cm$^{-1}$. These are followed by the quantum numbers of the upper energy levels, with the quantum numbers of the lower energy levels listed afterward. The final column consists of a line tag, which helps to track the source of each transition. This tag is constructed using the last two digits of the publication year, followed by the first two letters of each author's forename (up to eight letters in total), or the full surname if there is only one author. Each transition within a given source must be sequentially numbered to facilitate error detection. For example, a valid tag could be '15LiGoRo.1'. 

For the  CO isotopologues, only one vibrational ($v$) and one rotational ($J$) quantum numbers are considered. In the MARVEL input file, these quantum numbers are recorded for both the upper and lower energy levels.

All of the isotopologues except 28 exhibit hyperfine structure, the perturbation from the hyperfine structure is small and is only significant in lower energy transitions, mainly purely rotational transitions in the ground vibrational state. To use the these transitions for a MARVEL analysis we simply averaged over the wavenumbers of each hyperfine multiplet to give a transtion in terms of  $(v^\prime, J^\prime)$ and ($v^{\prime\prime},J^{\prime\prime}$).

\subsection{Quantum Numbers}
The MARVEL algorithm relies on uniquely labeled transitions to determine the empirical energy levels. Although MARVEL requires distinct quantum numbers for each state, it treats them as text labels.  For the CO molecule in its electronic ground state, each rovibrational energy level was labeled using  the vibrational quantum number ($v$) and the total angular momentum quantum number ($J$).

\section{Overview of the Literature}

High resolution rovibrational transition wavenumbers measured for CO isotopologues was extracted from 27 sources 
in the scientific literature and to construct MARVEL networks. A list of these sources is given in Table \ref{tab:LitTable}
which also gives information on the isotopologue(s) studied in each paper.
Tables~\ref{tab:sources27},  \ref{tab:sources28}, \ref{tab:sources36}, \ref{tab:sources37} and \ref{tab:sources38} summarize the final MARVEL run for 
each isotopologue. These tables show that we used 3731, 1461, 90, 728, 57 validated transitions for $^{13}$C$^{16}$O, $^{12}$C$^{18}$O,  $^{12}$C$^{17}$O,  $^{13}$C$^{18}$O and  $^{13}$C$^{17}$O, respectively.

\begin{table}
\caption{Summary of the experimental sources used to construct MARVEL networks. } \label{tab:LitTable} 
\begin{tabular}{lll} 
\hline \hline 
\multicolumn{1}{c}{Source tag} & \multicolumn{1}{c}{Reference} & \multicolumn{1}{c}{Isotopologue data}  \\ 
\hline   
58RoNeTo & \citet{58RoNeTo} & 36, 28, 38 \\ 
73Hall & \citet{73Hall} & 36,28,27 \\ 
74EnKiMiSp & \citet{74EnKiMiSp} & 36 \\
74JoMcWe & \citet{74JoMcWe} & 36 \\ 
76ChRaMc & \citet{76ChRaMc} & 36,28,38 \\ 
76ToClTeMc & \citet{76ToClTeMc} & 36, 28 \\ 
79Guelachvili & \citet{79Guelachvili} & 36,28,27,38,37 \\ 
79DaHeJoMc & \citet{79DaHeJoMc} & 36,28,38 \\
81FrLa & \citet{81FrLa}& 27\\
83GuDeFaUr & \citet{83GuDeFaUr} & 36,28 \\ 
85WiWiWi & \citet{85WiWiWi} & 36,28,38,37 \\ 
90ZiNaPaPr & \citet{90ZiNaPaPr} & 36 \\ 
90MaWeJe & \citet{90MaWeJe} & 36, 28,38 \\ 
91FaGuSaGr & \citet{91FaGuSaGr} & 36,28,27 \\ 
91NaInOrZi & \citet{91NaInOrZi} & 28 \\ 
94GeSaWaUr & \citet{94GeSaWaUr} & 36,28 \\ 
97WaSaMeGe & \citet{97WaSaMeGe} & 36,28 \\ 
00KlLeBeWi & \citet{00KlLeBeWi} & 38 \\
00KlLeGeBe & \citet{00KlLeGeBe} & 36 \\ 
01KlLeGeBe & \citet{01KlLeGeBe} & 28 \\ 
01BePaWoKl & \citet{01BePaWoKl} & 37 \\ 
02CaDoClPu & \citet{ 02CaDoClPu} & 36,27 \\ 
03PuDoCa & \citet{03PuDoCa} & 38,37 \\ 
03KlSuLeMu & \citet{03KlSuLeMu} & 27,37 \\ 
04CoHa & \citet{04CoHa} & 37\\
15CaKaKa & \citet{15CaKaKa} & 36,28,27 \\ 
15LiGoRo & \citet{15LiGoRo.CO} & 36,28,27 \\ 
\hline\hline
\end{tabular}
\end{table}

\subsection{Comments on the Data Sources (Table \ref{tab:LitTable})}

Data for 83GuDeFaUr, 91FaGuSaGr and 85WiWiWi could not be accessed or found, so they were extracted from  04CoHa which provides a  compilation of  data from many sources experimental studies of CO isotopologues. In each case the uncertainty for individual transitions was  set to be in the middle of the range 04CoHa provides. This approach did not cause any conflicts
in the MARVEL procedure.  Transitions cited as coming from 91FaGuSaGr and 83GuDeFaUr were not listed as a single source by 04CoHa, but are divided by the use of different filters and sets respectively. Here
we adopt the standard MARVEL convention of listing all transitions with common tag (91FaGuSaGr or 83GuDeFaUr). The precise location of these transitions in the 04CoHa data set is for \citet{91FaGuSaGr}:
\begin{itemize}
 \item[28]:
91FaGuSaGr.1 - 91FaGuSaGr.100 are Filter 2. 91FaGuSaGr.101 - 91FaGuSaGr.238 are Filter 3,
 \item[36]:
91FaGuSaGr.1 - 91FaGuSaGr.400 are Plate 1578. 91FaGuSaGr.401 - 91FaGuSaGr.681 are Filter 2. 91FaGuSaGr.682 - 91FaGuSaGr.1027 are Filter 3;
\end{itemize}
and for \citet{83GuDeFaUr}:
\begin{itemize}
 \item[28]:
All lines, 83GuDeFaUr.1 - 83GuDeFaUr.750 are Set A,
 \item[38]:
All lines, 83GuDeFaUr.1 - 83GuDeFaUr.536 are Set A,
 \item[36]:
83GuDeFaUr.1 - 83GuDeFaUr.760 are Set A. 83GuDeFaUr.761 - 83GuDeFaUr.1803 are Set B. 
\end{itemize}

Transitions measured by
94GeSaWaUr are included in the MARVEL input file, but was excluded from the calculations, as the experiment was later improved upon;  97WaSaMeGe gives improved accuracy for all  lines given by 94GeSaWaUr plus some additional transitions.

Some papers do not provide clearly stated uncertainties for transition wavenumbers, so the following values were used: for 15CaKaKa the uncertainty was set to 0.0005 cm$^{-1}$ which is the stated calibration error;
for 73Hall the uncertainty was set to 0.02 cm$^{-1}$ as this value makes the results consistent with other studies; for 15LiGoRo, the  uncertainty was set to 0.001 cm$^{-1}$ which is the stated  calibration error.

For the 27 isotopologue, the line Guelashvili.1 was removed, as it caused severe conflicts with other transitions in the MARVEL network and was therefore assumed to be incorrect.

\begin{center}
\begin{longtable}{llccccc} 
\caption{Experimental sources used to construct $^{12}$C$^{17}$O MARVEL network: Giving data sources, range of transition wavenumbers, A/V is the  number of measured(A) transitions and the number validated(V) by the MARVEL procedure, MSU is the average MARVEL suggested source uncertainty, CSU is average claimed source uncertainty, ASU is averaged uncertainty used in the final compilation. Range, MSU, CSU, ASU are all given in cm$^{-1}$.} \label{tab:sources27} \\ 
\hline \hline 
\multicolumn{1}{c}{Segment tag} & Reference& \multicolumn{1}{c}{Range} & \multicolumn{1}{c}{A/V} & \multicolumn{1}{c}{MSU}  & \multicolumn{1}{c}{CSU}  & \multicolumn{1}{c}{ASU}     \\ \hline  
\endfirsthead 
\multicolumn{3}{c} 
{{\bfseries \tablename \thetable{} -- continued from previous page}} \\ 
\multicolumn{1}{c}{Segment tag} & Reference& \multicolumn{1}{c}{Range} & \multicolumn{1}{c}{A/V} & \multicolumn{1}{c}{MSU}  & \multicolumn{1}{c}{CSU}  & \multicolumn{1}{c}{ASU}     \\ \hline  
\endhead 
\hline \multicolumn{3}{r}{{Continued on next page}} \\ \hline 
\endfoot 
\hline \hline 
\endlastfoot 
 02CaDoClPu & \citet{ 02CaDoClPu} & 3.74790232 - 14.99021366 & 4/4 & 3.336e-9 & 4.336e-8 & 2.001e-8  \\ 
85WiWiWi & \citet{85WiWiWi} & 3.7479023 - 14.9902150 & 4/4 & 8.350e-7 & 8.350e-7 & 8.350e-7  \\ 
03KlSuLeMu & \citet{03KlSuLeMu} & 3.7479013 - 29.9714962 & 7/7 & 1.330e-7 & 2.670e-6 & 9.536e-7  \\ 
81FrLa & \citet{81FrLa} & 3.747901  & 1/1 & 1.034e-6 & 1.034e-6 & 1.034e-6  \\ 
57RoNe & \citet{57RoNe} & 3.747902  & 1/1 & 2.001e-6 & 2.001e-6 & 2.001e-6  \\ 
79Guelachvili & \citet{79Guelachvili} & 1994.4967 - 2212.4431 & 55/54 & 1.100e-4 & 1.100e-4 & 1.100e-4  \\ 
15CaKaKa & \citet{15CaKaKa} & 8222.6293 - 8290.4854 & 6/6 & 5.000e-4 & 5.000e-4 & 5.000e-4  \\ 
15LiGoRo & \citet{15LiGoRo.CO} & 8290.4867 & 1/1 & 5.000e-4 & 5.000e-4 & 5.000e-4  \\ 
91FaGuSaGr & \citet{91FaGuSaGr} & 1874.690 - 1980.585 & 7/7 & 2.625e-3 & 2.625e-3 & 2.625e-3  \\ 
73Hall & \citet{73Hall} & 2141.57 - 2146.83 & 4/4 & 2.000e-2 & 2.000e-2 & 2.000e-2  \\ 
\end{longtable} 
\end{center}

\begin{center}
\begin{longtable}{llccccc} 
\caption{Experimental sources used to construct $^{12}$C$^{18}$O MARVEL network: Giving data sources, range of transition wavenumbers, A/V is the number of measured(A) transitions and the number validated(V) by the MARVEL procedure, MSU is the average MARVEL suggested source uncertainty, CSU is average claimed source uncertainty, ASU is averaged uncertainty used in the final compilation. Range, MSU, CSU, ASU are all given in cm$^{-1}$.} \label{tab:sources28} \\ 
\hline \hline 
\multicolumn{1}{c}{Segment tag} & Reference& \multicolumn{1}{c}{Range} & \multicolumn{1}{c}{A/V} & \multicolumn{1}{c}{MSU}  & \multicolumn{1}{c}{CSU}  & \multicolumn{1}{c}{ASU}     \\ \hline  
\endfirsthead 
\multicolumn{3}{c} 
{{\bfseries \tablename \thetable{} -- continued from previous page}} \\ 
\multicolumn{1}{c}{Segment tag} & Reference& \multicolumn{1}{c}{Range} & \multicolumn{1}{c}{A/V} & \multicolumn{1}{c}{MSU}  & \multicolumn{1}{c}{CSU}  & \multicolumn{1}{c}{ASU}     \\ \hline  
\endhead 
\hline \multicolumn{3}{r}{{Continued on next page}} \\ \hline 
\endfoot 
\hline \hline 
\endlastfoot 
01KlLeGeBe & \citet{01KlLeGeBe} & 3.66193909 - 65.78587344 & 11/11 & 3.340e-8 & 2.000e-7 & 7.733e-8  \\ 
58RoNeTo & \citet{58RoNeTo} & 3.6619394  & 1/1 & 2.670e-7 & 2.670e-7 & 2.670e-7  \\ 
85WiWiWi & \citet{85WiWiWi} & 3.6619391 - 18.3070308 & 5/5 & 8.350e-7 & 8.350e-7 & 8.350e-7  \\ 
94GeSaWaUr & \citet{94GeSaWaUr} & 2025.219903 - 2058.009052 & 7/0 & 5.670e-7 & 3.140e-6 & 1.106e-6  \\ 
97WaSaMeGe & \citet{97WaSaMeGe} & 1977.777505 - 2058.009052 & 11/11 & 5.340e-7 & 3.140e-6 & 1.152e-6  \\ 
91NaInOrZi & \citet{91NaInOrZi} & 32.94146 - 91.20244 & 17/17 & 4.000e-5 & 4.000e-5 & 4.000e-5  \\ 
79Guelachvili & \citet{79Guelachvili} & 1959.5310 - 2198.8463 & 64/64 & 1.000e-4 & 1.000e-4 & 1.000e-4  \\ 
15CaKaKa & \citet{15CaKaKa} & 8119.6393 - 8266.3400 & 38/38 & 5.000e-4 & 5.000e-4 & 5.000e-4  \\ 
90MaWeJe & \citet{90MaWeJe} & 2008.5943 - 2069.6666 & 2/2 & 5.000e-4 & 5.000e-4 & 5.000e-4  \\ 
76ToClTeMc & \citet{76ToClTeMc} & 1947.6469 - 2167.0995 & 122/122 & 7.000e-4 & 1.200e-3 & 8.131e-4  \\ 
15LiGoRo & \citet{15LiGoRo.CO} & 8206.047 & 1/1 & 1.000e-3 & 1.000e-3 & 1.000e-3  \\ 
91FaGuSaGr & \citet{91FaGuSaGr} & 1613.886 - 2269.557 & 238/238 & 2.000e-3 & 2.625e-3 & 2.362e-3  \\ 
83GuDeFaUr & \citet{83GuDeFaUr} & 1733.940 - 2203.624 & 750/750 & 2.710e-3 & 2.710e-3 & 2.710e-3  \\ 
79DaHeJoMc & \citet{79DaHeJoMc} & 1391.064 - 1649.905 & 87/87 & 5.000e-3 & 5.000e-3 & 5.000e-3  \\ 
76ChRaMc & \citet{76ChRaMc} & 1977.778 - 4213.692 & 95/95 & 5.000e-3 & 5.000e-3 & 5.000e-3  \\ 
73Hall & \citet{73Hall} & 2141.13 - 2146.45 & 12/12 & 2.000e-2 & 2.000e-2 & 2.000e-2  \\ 
\end{longtable} 
\end{center}

\begin{center}
\begin{longtable}{llccccc} 
\caption{Experimental sources used to construct $^{13}$C$^{16}$O MARVEL network: Giving data sources, range of transition wavenumbers, A/V is the number of measured(A) transitions and the number validated(V) by the MARVEL procedure, MSU is the average MARVEL suggested source uncertainty, CSU is average claimed source uncertainty, ASU is averaged uncertainty used in the final compilation. Range, MSU, CSU, ASU are all given in cm$^{-1}$.} \label{tab:sources36} \\ 
\hline \hline 
\multicolumn{1}{c}{Segment tag} & Reference& \multicolumn{1}{c}{Range} & \multicolumn{1}{c}{A/V} & \multicolumn{1}{c}{MSU}  & \multicolumn{1}{c}{CSU}  & \multicolumn{1}{c}{ASU}     \\ \hline  
\endfirsthead 
\multicolumn{3}{c} 
{{\bfseries \tablename \thetable{} -- continued from previous page}} \\ 
\multicolumn{1}{c}{Segment tag} & Reference& \multicolumn{1}{c}{Range} & \multicolumn{1}{c}{A/V} & \multicolumn{1}{c}{MSU}  & \multicolumn{1}{c}{CSU}  & \multicolumn{1}{c}{ASU}     \\ \hline  
\endhead 
\hline \multicolumn{3}{r}{{Continued on next page}} \\ \hline 
\endfoot 
\hline \hline 
\endlastfoot 
02CaDoClPu & \citet{02CaDoClPu} & 3.67592149 - 3.67592149 & 1/1 & 3.340e-8 & 3.340e-8 & 3.340e-8  \\ 
58RoNeTo & \citet{58RoNeTo} & 3.6759215 - 3.6759215 & 1/1 & 2.670e-7 & 2.670e-7 & 2.670e-7  \\ 
00KlLeGeBe & \citet{00KlLeGeBe} & 7.3517087 - 66.0365643 & 12/12 & 6.670e-8 & 3.340e-6 & 7.365e-7  \\ 
97WaSaMeGe & \citet{97WaSaMeGe} & 1994.702706 - 2065.754733 & 9/9 & 6.670e-7 & 3.900e-6 & 1.226e-6  \\ 
94GeSaWaUr & \citet{94GeSaWaUr} & 2028.907189 - 2065.754733 & 7/7 & 6.670e-7 & 3.900e-6 & 1.390e-6  \\ 
90ZiNaPaPr & \citet{90ZiNaPaPr} & 22.05083 - 109.67496 & 27/27 & 1.670e-6 & 3.340e-5 & 2.412e-5  \\ 
85WiWiWi & \citet{85WiWiWi} & 7.35171 - 18.37692 & 4/4 & 3.340e-5 & 3.340e-5 & 3.340e-5  \\ 
74EnKiMiSp & \citet{74EnKiMiSp} & 1877.6250 - 1877.6250 & 1/1 & 1.000e-4 & 1.000e-4 & 1.000e-4  \\ 
90MaWeJe & \citet{90MaWeJe} & 2025.0247 - 2069.6558 & 2/2 & 1.334e-4 & 3.336e-4 & 2.335e-4  \\ 
79Guelachvili & \citet{79Guelachvili} & 1948.9001 - 2210.2852 & 87/87 & 9.000e-5 & 1.700e-3 & 4.046e-4  \\ 
83GuDeFaUr & \citet{83GuDeFaUr} & 1205.5030 - 3060.8957 & 1803/1801 & 5.000e-4 & 5.000e-4 & 5.000e-4  \\ 
15CaKaKa & \citet{15CaKaKa} & 8094.9180 - 8273.5545 & 35/24 & 5.000e-4 & 5.000e-4 & 5.000e-4  \\ 
76ToClTeMc & \citet{76ToClTeMc} & 1831.6116 - 2185.2192 & 370/368 & 5.000e-4 & 3.200e-3 & 8.630e-4  \\ 
15LiGoRo & \citet{15LiGoRo.CO} & 8254.688 - 8254.688 & 1/1 & 1.000e-3 & 1.000e-3 & 1.000e-3  \\ 
91FaGuSaGr & \citet{91FaGuSaGr} & 1534.793 - 2276.305 & 1019/1019 & 8.000e-4 & 2.625e-3 & 1.751e-3  \\ 
74JoMcWe & \citet{74JoMcWe} & 1511.134 - 1885.626 & 160/160 & 3.000e-3 & 3.000e-3 & 3.000e-3  \\ 
79DaHeJoMc & \citet{79DaHeJoMc} & 1461.392 - 1514.494 & 19/19 & 5.000e-3 & 5.000e-3 & 5.000e-3  \\ 
76ChRaMc & \citet{76ChRaMc} & 1962.941 - 6266.639 & 162/162 & 5.000e-3 & 5.000e-3 & 5.000e-3  \\ 
73Hall & \citet{73Hall} & 2140.48 - 2145.25 & 11/11 & 2.000e-2 & 2.000e-2 & 2.000e-2  \\

\end{longtable} 
\end{center}

\begin{center}
\begin{longtable}{llccccc} 
\caption{Experimental sources used to construct $^{13}$C$^{17}$O MARVEL network: Giving data sources, range of transition wavenumbers, A/V is the number of measured(A) transitions and the number validated(V) by the MARVEL procedure, MSU is the average MARVEL suggested source uncertainty, CSU is average claimed source uncertainty, ASU is averaged uncertainty used in the final compilation. Range, MSU, CSU, ASU are all given in cm$^{-1}$.}\label{tab:sources37} \\ 
\hline \hline 
\multicolumn{1}{c}{Segment tag} & Reference& \multicolumn{1}{c}{Range} & \multicolumn{1}{c}{A/V} & \multicolumn{1}{c}{MSU}  & \multicolumn{1}{c}{CSU}  & \multicolumn{1}{c}{ASU}     \\ \hline  
\endfirsthead 
\multicolumn{3}{c} 
{{\bfseries \tablename \thetable{} -- continued from previous page}} \\ 
\multicolumn{1}{c}{Segment tag} & Reference& \multicolumn{1}{c}{Range} & \multicolumn{1}{c}{A/V} & \multicolumn{1}{c}{MSU}  & \multicolumn{1}{c}{CSU}  & \multicolumn{1}{c}{ASU}     \\ \hline  
\endhead 
\hline \multicolumn{3}{r}{{Continued on next page}} \\ \hline 
\endfoot 
\hline \hline 
\endlastfoot 
03PuDoCa & \citet{03PuDoCa} & 3.57877323 - 14.31382173 & 4/4 & 3.340e-8 & 6.670e-8 & 5.005e-8  \\ 
85WiWiWi & \citet{85WiWiWi} & 3.5787742 - 10.7358106 & 3/3 & 3.900e-7 & 6.670e-7 & 5.223e-7  \\ 
03KlSuLeMu & \citet{03KlSuLeMu} & 3.578761 - 32.193701 & 5/5 & 3.340e-7 & 3.340e-6 & 1.269e-6  \\ 
04CoHa$^a$ & \citet{04CoHa} & 10.735811 - 10.735811 & 1/1 & 3.000e-6 & 3.000e-6 & 3.000e-6  \\ 
01BePaWoKl & \citet{01BePaWoKl} & 3.57877 - 7.15742 & 2/2 & 1.510e-5 & 1.520e-5 & 1.515e-5  \\ 
79Guelachvili & \citet{79Guelachvili} & 1986.8945 - 2139.0096 & 43/43 & 3.000e-4 & 3.000e-4 & 3.000e-4  \\ 
\end{longtable} 

$^a$ An anonymous data source quoted by   \citet{04CoHa}.
\end{center}

\begin{center}
\begin{longtable}{llccccc}  
\caption{Experimental sources used to construct $^{13}$C$^{18}$O MARVEL network: Giving data sources, range of transition wavenumbers, A/V is the number of measured(A) transitions and the number validated(V) by the MARVEL procedure, MSU is the average MARVEL suggested source uncertainty, CSU is average claimed source uncertainty, ASU is averaged uncertainty used in the final compilation. Range, MSU, CSU, ASU are all given in cm$^{-1}$.}\label{tab:sources38} \\ 
\hline \hline 
\multicolumn{1}{c}{Segment tag} & Reference& \multicolumn{1}{c}{Range} & \multicolumn{1}{c}{A/V} & \multicolumn{1}{c}{MSU}  & \multicolumn{1}{c}{CSU}  & \multicolumn{1}{c}{ASU}     \\ \hline  
\endfirsthead 
\multicolumn{3}{c} 
{{\bfseries \tablename \thetable{} -- continued from previous page}} \\ 
\multicolumn{1}{c}{Segment tag} & Reference& \multicolumn{1}{c}{Range} & \multicolumn{1}{c}{A/V} & \multicolumn{1}{c}{MSU}  & \multicolumn{1}{c}{CSU}  & \multicolumn{1}{c}{ASU}     \\ \hline  
\endhead 
\hline \multicolumn{3}{r}{{Continued on next page}} \\ \hline 
\endfoot 
\hline \hline 
\endlastfoot 

03PuDoCa & \citet{03PuDoCa} & 3.49279596 - 6.98547125 & 2/2 & 6.671e-8 & 6.671e-8 & 6.671e-8  \\ 
58RoNeTo & \citet{58RoNeTo} & 3.4927969 - 3.4927969 & 1/1 & 2.670e-7 & 2.670e-7 & 2.670e-7  \\ 
00KlLeBeWi & \citet{00KlLeBeWi} & 6.9854711 - 31.4206269 & 8/8 & 1.334e-7 & 2.000e-6 & 8.334e-7  \\ 
85WiWiWi & \citet{85WiWiWi} & 3.4927965 - 17.4615573 & 5/5 & 8.350e-7 & 8.350e-7 & 8.350e-7  \\ 
90MaWeJe & \citet{90MaWeJe} & 2067.2860 - 2067.2860 & 1/1 & 5.003e-4 & 5.003e-4 & 5.003e-4  \\ 
79Guelachvili & \citet{79Guelachvili} & 1955.938 - 2126.149 & 43/43 & 2.000e-3 & 2.000e-3 & 2.000e-3  \\ 
83GuDeFaUr & \citet{83GuDeFaUr} & 1742.588 - 2131.313 & 536/536 & 2.710e-3 & 2.710e-3 & 2.710e-3  \\ 
76ChRaMc & \citet{76ChRaMc} & 1964.170 - 2109.858 & 40/40 & 5.000e-3 & 5.000e-3 & 5.000e-3  \\ 
79DaHeJoMc & \citet{79DaHeJoMc} & 1408.350 - 1626.718 & 92/92 & 5.000e-3 & 5.000e-3 & 5.000e-3  \\

\end{longtable} 
\end{center}

\section{Results and discussion} \label{sec:floats}

\subsection{MARVEL Energy Levels}

\begin{center}
\scriptsize
\begin{longtable}{cclcccc} 

\caption{Summary of results for different isotopologues, showing the number of measured (A) and validated (V) transitions, the number of energy levels
($N_{lev}$) output from MARVEL, natural abundance as given by HITRAN   \citep{jt836}, Acc. with for the number of energy levels generated from HITRAN which agree with the MARVEL ones within their stated  uncertainty;
data for $^{12}$C$^{16}$O is due to \citet{jt961}.} \label{tab:Comparison} \\ 
\hline \hline 
\multicolumn{1}{c}{Isotopologue} & Short Name& \multicolumn{1}{c}{Natural Abundance} & \multicolumn{1}{c}{Reduced Mass (amu)} & \multicolumn{1}{c}{A/V} & \multicolumn{1}{c}{$N_{lev}$} & \multicolumn{1}{c}{Acc. with HITRAN}  \\ 
\hline  
\endfirsthead 

\multicolumn{6}{c}{{\bfseries Table \thetable{} -- continued from previous page}} \\ 
\hline
\multicolumn{1}{c}{Isotopologue} & Short Name&\multicolumn{1}{c}{Natural Abundance} & \multicolumn{1}{c}{Reduced Mass (Da)} & \multicolumn{1}{c}{A/V} & \multicolumn{1}{c}{$N_{lev}$} & \multicolumn{1}{c}{Acc. with HITRAN}  \\ 
\hline  
\endhead 

\hline 
\multicolumn{6}{r}{{Continued on next page}} \\ 
\hline 
\endfoot 

\hline \hline 
\endlastfoot 

$^{12}$C$^{16}$O & 26 & 0.986544 & 6.8571 & 19399/19219 & 2293 & - \\ 
$^{13}$C$^{16}$O & 36 & 0.011084 & 7.1721 & 3731/3716 & 863 & 34/37 \\ 
$^{12}$C$^{18}$O & 28 & 0.001978 & 7.1997 & 1461/1454 & 499 & 35/35 \\ 
$^{12}$C$^{17}$O & 27 & 3.678670×10$^{-4}$ & 7.0000 & 90/89 & 33 & 9/9\\ 
$^{13}$C$^{18}$O & 38 & 2.222500×10$^{-5}$ & 7.5484 & 728/728 & 345 & 28/28\\ 
$^{13}$C$^{17}$O & 37 & 4.132920×10$^{-6}$ & 7.3682 & 57/57 & 45 & 21/21\\ 
\end{longtable}
\end{center}

Table~\ref{tab:Comparison} summarizes the results for the various isotopologues of CO including $^{12}$C$^{16}$O.
The table also gives the
natural abundance for each isotopologue which influences its ease of observation and hence how much data there is available. There is an obvious  correlation between the natural abundance and the number lines used/energy levels obtained for each
isotopologue. Unsurprisingly, there is much more information available for the parent 26 isotopologue, and the next  most abundant isotopologue, 36, yielded the most transitions and has the most energy levels among those species considered here.

The comparison with HITRAN given in Table \ref{tab:Comparison} tests whether our MARVEL energy levels reproduce HITRAN2020 \citep{jt836} transition wavenumbers  within their uncertainty. The comparison shows good agreement between
the HITRAN wavenumbers, which are all calculated by \citet{15LiGoRo.CO}, and our empirical MARVEL values.

\begin{figure}[h!]
    \centering
    \includegraphics[width=0.4\linewidth]{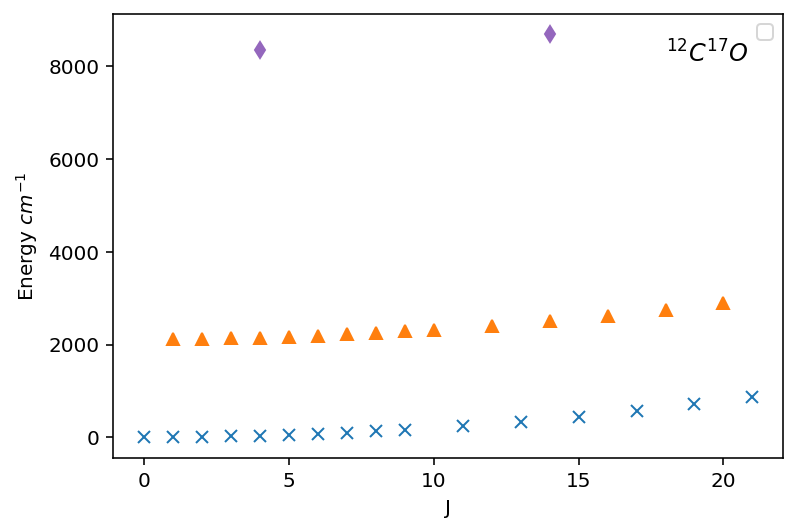}
       \includegraphics[width=0.4\linewidth]{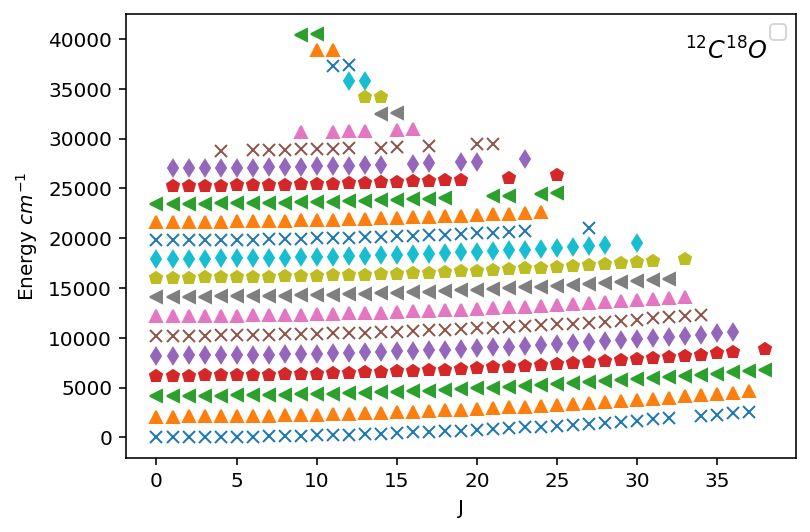}
          \includegraphics[width=0.4\linewidth]{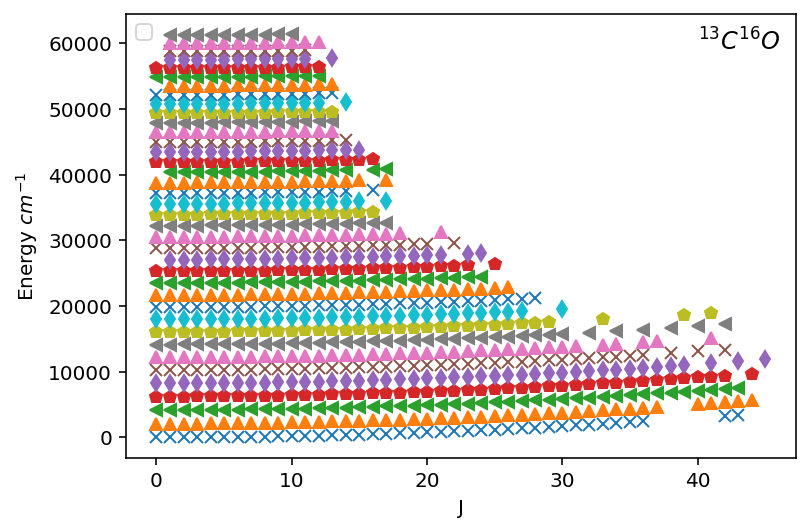}
             \includegraphics[width=0.4\linewidth]{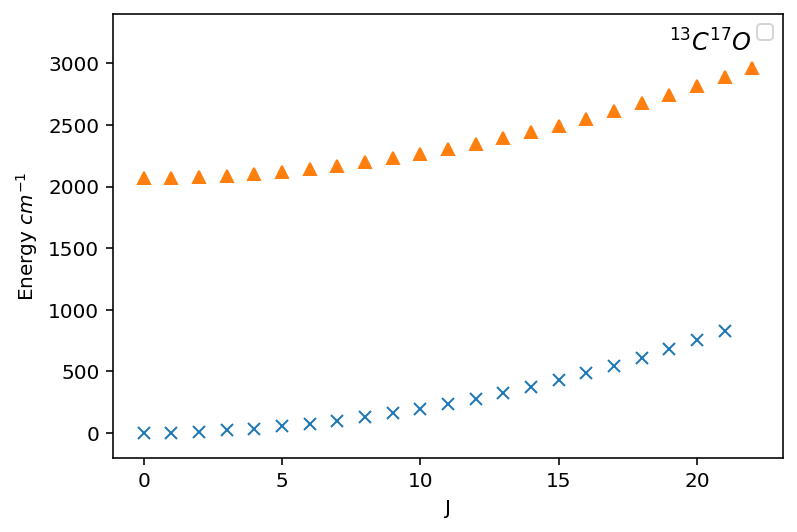}
                \includegraphics[width=0.4\linewidth]{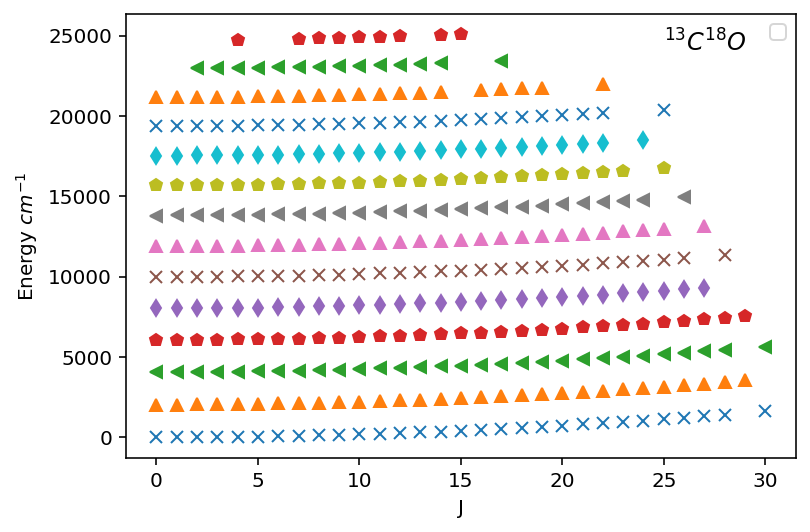}
    \caption{Ro-vibrational energy the CO isotopologues  as a function of $J$ rotational quantum number. }
    \label{fig:Jplot}
\end{figure}

Figure \ref{fig:Jplot} presents plots of the isotopologue energy levels as a function of $J$ quantum number for each of the five isotopologues. The vibrational states considered
for each isotopologue can be seen from the curves in figure; symbols and colors are used to help
distinguish them.

\section{ Summary and Conclusions }

We present accurate empirical energy levels for the five stable isotopologues of carbon monoxide. These levels are useful
for predicting accurate transition wavenumbers and for studies of isotopologue-dependent properties.
As demonstrated here, it is usual for many more energy levels to be determined by high resolution spectroscopy experiments for the parent molecule, here $^{12}$C$^{16}$O, than the less abundant, isotopically-substituted species. Isotopologue extrapolation methods have been proposed to predict the missing energy levels for minor isotopologues starting from empircal energy levels for the
parent species  \citep{jt665,jt948}.
These give reasonable results but sometimes need ad hoc corrections or are found not to work for well for certain bands,
see \citet{jt999} for example.
It seems likely that a machine learning approach to this problem will provide better results. The current dataset of energy levels for 6 isotopologues provides a useful starting point for such an approach \citep{jtMLiso}.

We note that MARVEL energies are consistent with those that can obtained  from HITRAN within our MARVEL uncertainties. This was
also largely true for the parent $^{12}$C$^{16}$O isotopologue \citep{jt961}, the exception being for states with a high degree of both vibrational and rotational excitation; experimental transitions involving  the corresponding isotopologue levels are currently not available. This means that improvements may be needed for
analysis of  CO isotopologies spectra in hot environments, such as solar spectra, but not for the analysis of cooler, interstellar environments.

However, our empirical energy levels/wavenumber are likely more accurate than those given by HITRAN and possess individual uncertainties. Recently, CO has become the focus of a variety of ultrahigh accuracy studies \citep{25LiDaWo.CO,jt987,25HuWaYi.CO}; it is likely that these studies will be extended to consider isotopologue ratios in the near future, see \citet{21NaturePhys.CO2} for example. The accurate isotopologue results presented here will form an important component of theoretical models that have been used to underpin these studies \citep{jt871,jt970}.

\section{Acknowledgments}

\begin{acknowledgments}
This work was supported by the STFC grant UKRI/ST/B001183/1.
\end{acknowledgments}

%




\appendix 
\section{MARVEL files}

The MARVEL input and output files are given in machine readable form
in the Supplementary Material.
Tables~\ref{table:trans27}, \ref{table:trans28}, \ref{table:trans36}, \ref{table:trans37} and \ref{table:trans38}
contain extracts of the input transitions for each isotopologue.
Tables \ref{table:energies27}, \ref{table:energies28}, \ref{table:energies36}, \ref{table:energies37} and \ref{table:energies38} contain extracts of the respective output energies files.

\clearpage

\begin{table}
\caption{MARVEL transitions  file for $^{12}$C$^{17}$O (27).}
\label{table:trans27}
\begin{tabular}{rcrrrrl}
\hline\hline
 $\nu$/cm$^{-1}$~~~ & $u$/cm$^{-1}$  & 
 $v^\prime$ & $J^\prime$ & $v^{\prime\prime}$ & $J^{\prime\prime}$ & tag\\ 
 \hline
3.747901334&	1.67E-06&	0&	1&	0&	0&	03KlSuLeMu.1\\
7.495674499&	2.67E-06&	0&	2&	0&	1&	03KlSuLeMu.2\\
11.243146&	1.33E-07&	0&	3&	0&	2&	03KlSuLeMu.3\\
14.99021476&	5.34E-07&	0&	4&	0&	3&	03KlSuLeMu.4\\
18.73672001&	6.67E-07&	0&	5&	0&	4&	03KlSuLeMu.5\\
26.22750135&	6.67E-07&	0&	7&	0&	6&	03KlSuLeMu.6\\
29.97149624&	3.34E-07&	0&	8&	0&	7&	03KlSuLeMu.7\\  
 \hline
\end{tabular}

\vspace{0.2cm}  
\textbf{Note:}
$\nu$ is the wavenumber entry for the transition,
$u$ is the associated uncertainty
of the transition, $v$ and $J$ are the vibrational and rotational
quantum numbers, respectively, of the initial ($''$) and final ($'$) states, 
while tag refers to the source of the transition.\\
(This table is available in its entirety in machine-readable form in the online article.)
\end{table}

\begin{table}
\caption{MARVEL transitions  file for $^{12}$C$^{18}$O (28).}
\label{table:trans28}
\begin{tabular}{rcrrrrl}
\hline\hline
 $\nu$/cm$^{-1}$~~~ & $u$/cm$^{-1}$  & 
 $v^\prime$ & $J^\prime$ & $v^{\prime\prime}$ & $J^{\prime\prime}$ & tag\\ 
 \hline
32.94146246 &    4.00E-05 &       0   &    9    &   0  &     8    &   91NaInOrZi.1   \\
36.59743502 &    4.00E-05 &       0   &    10   &   0  &     9    &   91NaInOrZi.2   \\
40.25200327 &    4.00E-05 &       0   &    11   &   0  &     10   &   91NaInOrZi.3   \\
43.90518723 &    4.00E-05 &       0   &    12   &   0  &     11   &   91NaInOrZi.4   \\
47.55674674 &    4.00E-05 &       0   &    13   &   0  &     12   &   91NaInOrZi.5   \\
51.20653502 &    4.00E-05 &       0   &    14   &   0  &     13   &   91NaInOrZi.6   \\
54.85452539 &    4.00E-05 &       0   &    15   &   0  &     14   &   91NaInOrZi.7   \\  
 \hline
\end{tabular}

\vspace{0.2cm}  
\textbf{Note:}
$\nu$ is the wavenumber entry for the transition,
$u$ is the associated uncertainty
of the transition, $v$ and $J$ are the vibrational and rotational
quantum numbers, respectively, of the initial ($''$) and final ($'$) states, 
while tag refers to the source of the transition.\\
(This table is available in its entirety in machine-readable form in the online article.)
\end{table}

\begin{table}
\caption{MARVEL transitions  file for $^{13}$C$^{16}$O (36).}
\label{table:trans36}
\begin{tabular}{rcrrrrl}
\hline\hline
 $\nu$/cm$^{-1}$~~~ & $u$/cm$^{-1}$  & 
 $v^\prime$ & $J^\prime$ & $v^{\prime\prime}$ & $J^{\prime\prime}$ & tag\\ 
 \hline
2065.754733  &   7.00E-07   &     1   &    7     &  0   &    8 &      94GeSaWaUr.1  \\
2061.821909  &   7.00E-07   &     1   &    8     &  0   &    9 &      94GeSaWaUr.2  \\
2057.857576  &   8.01E-07   &     1   &    9     &  0   &    10&      94GeSaWaUr.3  \\
2053.861869  &   6.67E-07   &     1   &    10    &  0   &    11&      94GeSaWaUr.4  \\
2045.776869  &   2.23E-06   &     1   &    12    &  0   &    13&      94GeSaWaUr.5  \\
2041.687842  &   7.34E-07   &     1   &    13    &  0   &    14&      94GeSaWaUr.6  \\
2028.907189  &   3.90E-06   &     2   &    10    &  1   &    11&      94GeSaWaUr.7  \\
 \hline
\end{tabular}

\vspace{0.2cm}  
\textbf{Note:}
$\nu$ is the wavenumber entry for the transition,
$u$ is the associated uncertainty
of the transition, $v$ and $J$ are the vibrational and rotational
quantum numbers, respectively, of the initial ($''$) and final ($'$) states, 
while tag refers to the source of the transition.\\
(This table is available in its entirety in machine-readable form in the online article.)
\end{table}

\begin{table}
\caption{MARVEL transitions  file for $^{13}$C$^{17}$O (37).}
\label{table:trans37}
\begin{tabular}{rcrrrrl}
\hline\hline
 $\nu$/cm$^{-1}$~~~ & $u$/cm$^{-1}$  & 
 $v^\prime$ & $J^\prime$ & $v^{\prime\prime}$ & $J^{\prime\prime}$ & tag\\ 
 \hline
 3.578774237    &  5.10E-07      &   0     &   1    &    0    &    0     &   85WiWiWi.1        \\
 7.157420925    &  3.90E-07      &   0     &   2    &    0    &    1     &   85WiWiWi.2        \\
 10.73581056    &  6.67E-07      &   0     &   3    &    0    &    2     &   85WiWiWi.3        \\
 3.578760811    &  1.67E-06      &   0     &   1    &    0    &    0     &   03KlSuLeMu.1      \\
 7.15741398     &  6.67E-07      &   0     &   2    &    0    &    1     &   03KlSuLeMu.2      \\
 10.73581044    &  3.34E-06      &   0     &   3    &    0    &    2     &   03KlSuLeMu.3      \\
 28.61950597    &  3.34E-07      &   0     &   8    &    0    &    7     &   03KlSuLeMu.4      \\  
 \hline
\end{tabular}

\vspace{0.2cm}  
\textbf{Note:}
$\nu$ is the wavenumber entry for the transition,
$u$ is the associated uncertainty
of the transition, $v$ and $J$ are the vibrational and rotational
quantum numbers, respectively, of the initial ($''$) and final ($'$) states, 
while tag refers to the source of the transition.\\
(This table is available in its entirety in machine-readable form in the online article.)
\end{table}

\begin{table}
\caption{MARVEL transitions  file for $^{13}$C$^{18}$O (38).}
\label{table:trans38}
\begin{tabular}{rcrrrrl}
\hline\hline
 $\nu$/cm$^{-1}$~~~ & $u$/cm$^{-1}$  & 
 $v^\prime$ & $J^\prime$ & $v^{\prime\prime}$ & $J^{\prime\prime}$ & tag\\ 
 \hline
2040.19944 &2.000e-3& 1 &0 & 0 &1 &   79Guelachvili.1    \\ 
2036.67632 &2.000e-3& 1 &1 & 0 &2 &   79Guelachvili.2     \\
2033.12345 &2.000e-3& 1 &2 & 0 &3 &   79Guelachvili.3     \\
2029.54036 &2.000e-3& 1 &3 & 0 &4 &   79Guelachvili.4     \\
2025.92767 &2.000e-3& 1 &4 & 0 &5 &   79Guelachvili.5     \\
2022.2852  &2.000e-3& 1 &5 & 0 &6  &  79Guelachvili.6      \\
2018.61312 &2.000e-3& 1 &6 & 0 &7 &   79Guelachvili.7     \\
 \hline
\end{tabular}

\vspace{0.2cm}  
\textbf{Note:}
$\nu$ is the wavenumber entry for the transition,
$u$ is the associated uncertainty
of the transition, $v$ and $J$ are the vibrational and rotational
quantum numbers, respectively, of the initial ($''$) and final ($'$) states, 
while tag refers to the source of the transition.\\
(This table is available in its entirety in machine-readable form in the online article.)
\end{table}

\begin{table}
\caption{MARVEL energy levels  file  for $^{12}$C$^{17}$O (27).}
\label{table:energies27}
\begin{tabular}{rrrcr} 
 \hline\hline
 $v$ & $J$ & $E$/cm$^{-1}$~~~~& uncertainty/cm$^{-1}$ & $N$ \\ 
 \hline
0& 0& 	0.000000000&	0.000000000&	6\\
0& 1& 	3.747902324&	0.000002001&	8\\
0& 2&	11.243567407&	0.000002836&	8\\
0& 3& 	22.486716056&	0.000003671&	8\\
0& 4& 	37.476929727&	0.000004506&	6\\
0& 5& 	56.213649736&	0.000005173&	5\\
0& 6& 	78.696214049&	0.000224506&	4\\
 \hline
\end{tabular}

\vspace{0.2cm} 
\textbf{Note:} 
$v$ and $J$ are the vibrational and rotational
quantum numbers, respectively, $E$ is the empirical rovibrational energy level
determined through a MARVEL analysis, while $N$ is the number of incident
transitions.
\\
(This table is available in its entirety in machine-readable form in the online article.)
\end{table}

\begin{table}
\caption{MARVEL energy levels  file  for $^{12}$C$^{18}$O (28).}
\label{table:energies28}
\begin{tabular}{rrrcr} 
 \hline\hline
 $v$ & $J$ & $E$/cm$^{-1}$~~~~& uncertainty/cm$^{-1}$ & $N$ \\ 
 \hline
             0&  0  &    0.00000000   &    0.00000000   &    9      \\
              0&  1  &    3.66193911   &    0.00000084   &    15     \\
              0&  2  &    10.98568419  &    0.00000167   &    16     \\
              0&  3  &    21.97096897  &    0.00000251   &    17     \\
              0&  4  &    36.61739366  &    0.00000334   &    17     \\
              0&  5  &    54.92442544  &    0.00000417   &    16     \\
              0&  6  &    76.89139828  &    0.00000421   &    11     \\
 \hline
\end{tabular}

\vspace{0.2cm} 
\textbf{Note:} 
$v$ and $J$ are the vibrational and rotational
quantum numbers, respectively, $E$ is the empirical rovibrational energy level
determined through a MARVEL analysis, while $N$ is the number of incident
transitions.
\\
(This table is available in its entirety in machine-readable form in the online article.)
\end{table}

\begin{table}
\caption{MARVEL energy levels  file  for $^{13}$C$^{16}$O (36).}
\label{table:energies36}
\begin{tabular}{rrrcr} 
 \hline\hline
 $v$ & $J$ & $E$/cm$^{-1}$~~~~& uncertainty/cm$^{-1}$ & $N$ \\ 
 \hline
    0& 0   &  0.00000000    &  0.00000000  &    10         \\
    0& 1   &  3.67592149    &  0.00000003  &    21         \\
    0& 2   &  11.02763023   &  0.00000019  &    21         \\
    0& 3   &  22.05485775   &  0.00000026  &    22         \\
    0& 4   &  36.75720134   &  0.00000032  &    23         \\
    0& 5   &  55.13412408   &  0.00000039  &    21         \\
    0& 6   &  77.18495484   &  0.00000046  &    24         \\
 \hline
\end{tabular}

\vspace{0.2cm} 
\textbf{Note:} 
$v$ and $J$ are the vibrational and rotational
quantum numbers, respectively, $E$ is the empirical rovibrational energy level
determined through a MARVEL analysis, while $N$ is the number of incident
transitions.
\\
(This table is available in its entirety in machine-readable form in the online article.)
\end{table}

\begin{table}
\caption{MARVEL energy levels  file  for $^{13}$C$^{17}$O (37).}
\label{table:energies37}
\begin{tabular}{rrrcr} 
 \hline\hline
 $v$ & $J$ & $E$/cm$^{-1}$~~~~& uncertainty/cm$^{-1}$ & $N$ \\ 
 \hline
     0 &0   & 0.00000000   &   0.00000000  &    5    \\ 
     0 &1   &  3.57877323   &   0.00001510  &    9    \\
     0 &2   &  10.73619238  &   0.00003030  &    9    \\
     0 &3   &  21.47201365  &   0.00003330  &    7    \\
     0 &4   &  35.78583538  &   0.00003337  &    3    \\
     0 &5   &  53.67720514  &   0.00063330  &    2    \\
     0 &6   &  75.14539399  &   0.00063337  &    2    \\
 \hline
\end{tabular}

\vspace{0.2cm} 
\textbf{Note:} 
$v$ and $J$ are the vibrational and rotational
quantum numbers, respectively, $E$ is the empirical rovibrational energy level
determined through a MARVEL analysis, while $N$ is the number of incident
transitions.
\\
(This table is available in its entirety in machine-readable form in the online article.)
\end{table}

\begin{table}
\caption{MARVEL energy levels  file  for $^{13}$C$^{18}$O (38).}
\label{table:energies38}
\begin{tabular}{rrrcr} 
 \hline\hline
 $v$ & $J$ & $E$/cm$^{-1}$~~~~& uncertainty/cm$^{-1}$ & $N$ \\ 
 \hline
0& 0 &    0.00000000    &  0.00000000  &    6       \\  
0& 1 &    3.49279601    &  0.00000084  &    12       \\
0& 2 &    10.47826723   &  0.00000167  &    10       \\
0& 3 &    20.95617125   &  0.00000251  &    10       \\
0& 4 &    34.92614440   &  0.00000334  &    10       \\
0& 5 &    52.38770211   &  0.00000417  &    9        \\
0& 6 &    73.34023883   &  0.00000431  &    9        \\
 \hline
\end{tabular}

\vspace{0.2cm} 
\textbf{Note:} 
$v$ and $J$ are the vibrational and rotational
quantum numbers, respectively, $E$ is the empirical rovibrational energy level
determined through a MARVEL analysis, while $N$ is the number of incident
transitions.
\\
(This table is available in its entirety in machine-readable form in the online article.)
\end{table}

\clearpage

\bibliographystyle{aasjournal}



\end{document}